\documentstyle[11pt,paspconf,epsf]{article}

\begin{document}

 \title{Simultaneous   age-metallicity  estimates of  the Hyades  open
 cluster from three binary systems}

\author{E. Lastennet}
 \affil{Astronomy Unit,  Queen  Mary and  Westfield  College, Mile End
 Road, London E1 4NS, UK}
    
\author{D. Valls-Gabaud}
 \affil{UMR  CNRS   7550, Observatoire    Astronomique,  11,  rue   de
 l'Universit\'e, 67000 Strasbourg, France}
    
\author{Th. Lejeune}
 \affil{Astronomisches Institut der  Universit\"at Basel, Venusstr. 7,
 CH-4102 Binningen, Switzerland}
    
\author{E. Oblak}
 \affil{Observatoire de Besan\c  con, 41 bis avenue de l'Observatoire,
 F-25010 Besan\c con, Cedex, France}

\begin{abstract}
Three binary systems in the Hyades open cluster (51 Tau, V818 Tau, and
$\theta^2$ Tau), with known metallicity  and good Johnson  photometric
data  are used  to  test  the  validity of  three  independent sets of
stellar   evolutionary tracks. A  statistical  method is described and
applied  to   the  colour-magnitude  diagram   of the   six   selected
components,  giving rise to  $\chi^2$-contours in  the age-metallicity
plane. The effects of   the Hipparcos parallaxes on  these  confidence
regions are studied in detail  for these binaries through a comparison
with very  accurate   but   older  orbital parallaxes.     Independent
simultaneous  age-metallicity estimates  are  given and  compared with
observational constraints.
\end{abstract}


\keywords{binary systems, open cluster: Hyades}

\section{Introduction}

Since the members of an open cluster are assumed to be of same age and
chemical composition,  these  stars are currently   used  to test  the
validity of stellar evolution theories,  mainly because main  sequence
stars  define    a tight  sequence  in   a   colour-magnitude  diagram
(CMD). Unfortunately, this  tightness is sometimes  misleading because
of the  contamination   by field stars,    the presence of  unresolved
binaries and also the influence of stellar rotation on the location of
massive stars   in  CMDs. Alternatively, well-detached   binaries  are
powerful tests when  fundamental parameters are accurately known  (see
the  comprehensive review by  Andersen  1991 on double-lined eclipsing
binaries).    Unfortunately,  the determination    of   their chemical
composition often remains a difficult and unresolved issue. It appears
therefore that  a  better test  could be  performed by  combining both
advantages,   that    is,   testing the    tracks   with well-detached
double-lined binaries   which are members  of  open clusters.  We have
applied this idea to   three  well-detached binaries members  of   the
Hyades: 51 Tau, V818 Tau, and $\theta^2$ Tau.

\subsubsection{Observational data : } 
Torres et  al.,  1997 ([TSL97a], [TSL97b]  and  [TSL97c]) obtained the
first    complete   visual-spectroscopic     solutions   for  the    3
above-mention\-ned systems,  from  which  they carefully derived  very
accurate parallaxes and individual  masses.  They also gathered   some
individual photometric data  in  the Johnson system.  Furthermore,  we
found useful trigonometric   parallaxes information  in the  Hipparcos
catalogue (ESA, 1997).  By combining   the two  sources of data,    we
investigate  the influence of the  Hipparcos parallaxes  on our method
which  was  developed  to test   stellar  evolutionary  models  in  HR
diagrams.

\subsubsection{Theoretical tracks : }
Among  the   most widely used    stellar   theoretical tracks  in  the
literature  are those computed by  the Geneva group (see Charbonnel et
al. 1993 and references therein) and the Padova  group (see Fagotto et
al. 1994 and references therein). We also used the stellar tracks from
Claret \& Gim\'enez (1992) (CG92  thereafter). The tests are done with
these 3 series of stellar tracks.

\subsubsection{Tests in the CMD : }
The tests we want to perform are the following :
\begin{enumerate}
\item 
to  check whether the two  components of the  systems  are on the same
isochrone, i.e. on   a  line defined by   the  same age and  the  same
chemical composition for the two single stars.
\item 
since all the  selected    stars are  members  of the    Hyades  whose
metallicity  has   been  well measured (according   to  the  review of
Perryman  et al.    (1998):  [Fe/H]  $=$ 0.14   $\pm$    0.05, i.e:  Z
$=$$0.024^{+0.0025}_{-0.003}$), we  can also  check that the predicted
metallicities from theoretical models are correct.
\item 
for 51 Tau and $\theta^2$ Tau, the individual stellar masses are known
with an accuracy of about 10\%, and for V818 Tau, masses and radii are
known with an accuracy close to 1-2\%, allowing further tests with the
theoretical models.
\end{enumerate}

Therefore, if one  of these criteria  is  not clearly fullfilled  by a
given set  of  tracks, then these  models have  obvious problems since
they do not account for several  observational constraints (namely the
metallicity, mass, radius, and/or the photometric data).

\subsubsection{Photometric calibrations : } 
We do  not claim that  the 6 selected  Hyades  stars allow  us to test
without ambiguity  any set  of  theoretical stellar  tracks. Since the
data are   presented  in CMD, we  are   in fact testing   not only the
validity  of the tracks but also  of the photometric calibrations, and
disentangling the relative influence of both is a  tricky task. We use
the Basel     Stellar    Library (BaSeL)   photometric   calibrations,
extensively   tested and  regularly  updated   for   a larger set   of
parameters   (see   Lejeune  et  al.  1997,    1998  and  Lastennet et
al.  1999a). For  reasons developed in   Lastennet et al.  (1999b), we
assume that the calibrations from the BaSeL models are reliable enough
for this work (for more  details and references  on the BaSeL library,
see   contributions of Lejeune   et  al. and  Westera  et al.  in this
volume).

\subsubsection{Brief description of the statistical method : } 
In order to derive simultaneously the metallicity  (Z) and the age (t)
of the system,  and to produce confidence  level  contours (see Figure
1), we minimize the $\chi^2$-functional defined as:

\small
\begin{eqnarray}
 \chi^2 (t, Z) & = & \sum_{i=A}^{B} \left[ \left(\frac{\rm M_V(i)_{\rm
 mod}   - M_V(i)}   {\sigma(\rm  M_V(i))}\right)^2 +   \left(\frac{\rm
 (B-V)(i)_{\rm mod} - (B-V)(i)}{\sigma(\rm (B-V)(i))}\right)^2 \right]
\end{eqnarray}
\normalsize

where $A$ is the  primary and $B$ the  secondary component.  M$_V$ and
(B$-$V)  are  the   observed  values,   and  M$_V$$_{\rm  mod}$    and
(B$-$V)$_{\rm mod}$ are  obtained  from the synthetic  computations of
the BaSeL models using a given set of stellar tracks.
\begin{figure}
\plotfiddle{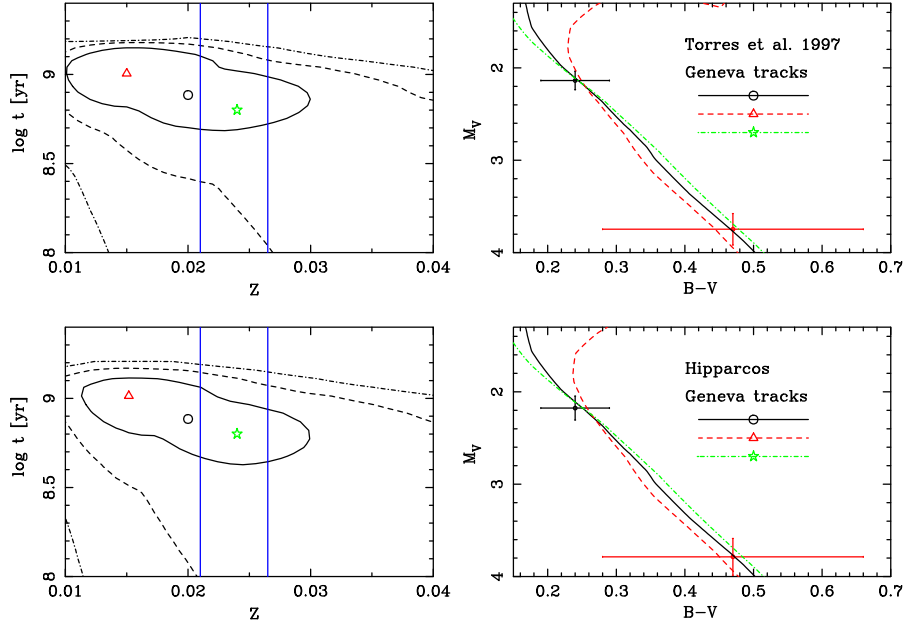}{8. cm}{-90}{45}{45}{-180}{240}
 \caption{51 Tau system  : influence of  the Hipparcos parallax on the
 contour levels derived from the Geneva tracks.   The location of each
 star in the CMD  is from Torres et  al. [TSL97a] (in the  {\it bottom
 panel}, $M_V$   is   derived  from  the  Hipparcos  parallax).     In
 isocontours plots, the best fits  ($\chi^2_{\rm min}$) are marked  by
 {\it open circles}.  The result of Perryman et al. (log t $=$ 8.80, Z
 $=$ 0.024)  is shown  for comparison ({\it   star}).  The 1,  2,  and
 3$\sigma$ contour levels  (respectively {\it solid}, {\it dashed} and
 {\it dot-dashed   lines}) are not  significantly  modified.  Vertical
 lines in contour  diagrams   show the  observational limits  for  the
 metallicity of the Hyades.} \label{fig-1}
\end{figure}
\normalsize

\section{Results}
Table below briefly summarizes the results (see Lastennet et al. 1999b
for further details)  of the theoretical simultaneous age--metallicity
estimates obtained from isochrone age fitting (1$\sigma$ level) taking
into account the Hipparcos parallax.

%

\begin{table}
\small
\begin{tabular}{lllllll}
\tableline 
\noalign{\smallskip} 
System         & \multicolumn{2}{c}{Geneva} & \multicolumn{2}{c}{Padova} & \multicolumn{2}{c}{CG92} \\
\noalign{\smallskip}
\tableline
\noalign{\smallskip}
               & Z & log t                  & Z & log t                   & Z & log t               \\ 
\noalign{\smallskip} 
\tableline
\noalign{\smallskip}
\noalign{\smallskip}
51 Tau         & 0.020$^{+0.010}_{-0.008}$ & 8.88$^{+0.22}_{-0.23}$ & 
0.017$^{+0.021}_{-0.005}$ & 8.90$^{+0.15}_{-0.55}$ & 
0.018$^{+0.012}_{-0.006}$ & 8.92$^{+0.23}_{-0.17}$ \\ 
\noalign{\smallskip}
V818 Tau
&                           &                        & 
0.033$^{+0.017}_{-0.015}$ & 7.30$^{+2.50}_{-0.30}$ & 
                          &                        \\ 
\noalign{\smallskip}
$\theta^2$ Tau & 0.027$^{+0.013}_{-0.010}$ & 8.80$^{+0.05}_{-0.09}$ & 
0.027$^{+0.023}_{-0.011}$ & 8.80$^{+0.03}_{-0.11}$ &
0.027$^{+0.003}_{-0.005}$ & 8.88$^{+0.02}_{-0.02}$ \\ 
\noalign{\smallskip}
\tableline
\tableline
\noalign{\smallskip} 
\end{tabular}
\end{table}

\begin{itemize}
\item 
For 51 Tau and $\theta^2$ Tau, the 3 sets of isochrones give good fits
in the CMD, in agreement with previous estimates  (Perryman et al.) of
age    (log  t  $=$   $8.80^{+0.02}_{-0.04}$,  from isochrone  fitting
technique with the CESAM  stellar evolutionary code (Morel  1997)) and
metallicity ([Fe/H] $=$ 0.14 $\pm$ 0.05).
\item 
The Geneva and CG92  models can not be  tested  with the less  massive
component of V818 Tau. The Padova tracks provide contours in agreement
with the Hyades  metallicity    only when  taking into    account  the
Hipparcos parallax. Otherwise, solutions are too old and metal rich.
\item 
Masses predicted by  the 3 sets of tracks  are in good agreement  with
the measured individual masses of each system.
\item 
Padova  isochrones can not fit  the system V818  Tau  in a mass-radius
diagram.
\end{itemize}


\end{document}